\documentclass[trackchanges]{aastex7}

\usepackage{physymb}
\usepackage{mathrsfs}
\usepackage{soul}
\usepackage{color}

\def\fexxv{Fe\,{\sc xxv}}
\def\fexxvi{Fe\,{\sc xxvi}}

\def\cm{\ifmmode {\rm cm}^{-1} \else cm$^{-1}$ \fi}
\def\s{\ifmmode {\rm s}^{-1} \else s$^{-1}$ \fi}
\def\cc{\ifmmode {\rm cm}^{-3} \else cm$^{-3}$ \fi}
\def\cs{\ifmmode {\rm cm}^{-2} \else cm$^{-2}$ \fi}
\def\g{\ifmmode \gamma \else $\gamma$\fi}
\def\G{\ifmmode \Gamma \else $\Gamma$\fi}
\def\Gs{\ifmmode \Gamma~ \else $\Gamma~$\fi}

\def\gc{\ifmmode \gamma_{\rm c} \else $\gamma_{\rm c}$ \fi}
\def\sw{Schwarzschild~}
\def\gsim{\mathrel{\raise.5ex\hbox{$>$}\mkern-14mu
             \lower0.6ex\hbox{$\sim$}}}
\def\lsim{\mathrel{\raise.3ex\hbox{$<$}\mkern-14mu
             \lower0.6ex\hbox{$\sim$}}}
\def\simless{\mathbin{\lower 3pt\hbox
     {$\rlap{\raise 5pt\hbox{$\char'074$}}\mathchar"7218$}}}   
\def\simmore{\mathbin{\lower 3pt\hbox
     {$\rlap{\raise 5pt\hbox{$\char'076$}}\mathchar"7218$}}}   
\def\Msun{M_\odot}                                
\def\deg{^\circ}

\def\gro1655{GRO~J1655-40}
\def\4u1630{4U1630-472}
\def\h1743{H1743-322}
\def\grs1915{GRS1915+105}

\begin{document}

\title{Probing Accretion Disk Winds of Stratified Nature with \fexxvi\ Doublet in Black Hole X-ray Binaries}

\author[orcid=0000-0001-5709-7606]{Keigo Fukumura}
\affiliation{Department of Physics and Astronomy, James Madison University, Harrisonburg, VA 22807, USA}
\email[show]{fukumukx@jmu.edu}  

\author[orcid=0000-0002-5701-0811]{Shoji Ogawa} 
\affiliation{Institute of Space and Astronautical Science (ISAS), Japan Aerospace Exploration Agency (JAXA), Kanagawa 252-5210, Japan}
\email{ogawa.shohji@jaxa.jp}

\author[orcid=0000-0002-0114-5581]{Atsushi Tanimoto} 
\affiliation{Graduate School of Science and Engineering, Kagoshima University, Kagoshima, 890-8580, Japan}
\email{atsushi.tanimoto@sci.kagoshima-u.ac.jp}

\author[orcid=0000-0002-6562-8654]{Francesco Tombesi} 
\affiliation{Physics Department, Tor Vergata University of Rome, Via della Ricerca Scientifica 1, 00133 Rome, Italy}
\affiliation{INAF—Astronomical Observatory of Rome, Via Frascati 33, 00040 Monte Porzio Catone, Italy}
\affiliation{INFN—Rome Tor Vergata, Via della Ricerca Scientifica 1, 00133 Rome, Italy}
\email{francesco.tombesi@roma2.infn.it}

\author[orcid=0000-0002-1035-8618]{Alfredo Luminari} 
\affiliation{INAF—IAPS, via del Fosso del Cavaliere 100, 00100 Roma, Italy}
\affiliation{INAF—Osservatorio Astronomico di Roma, Via Frascati 33, 00078 Monteporzio, Italy}
\email{alfredo.luminari@inaf.it}

\author[orcid=0009-0003-8610-853X]{Maxime Parra} 
\affiliation{Department of Physics, Ehime University, 2-5, Bunkyocho, Matsuyama,
Ehime 790-8577, Japan}
\email{maximeparrastro@gmail.com}

\author[orcid=0000-0001-8195-6546]{Megumi Shidatsu} 
\affiliation{Department of Physics, Ehime University, 2-5, Bunkyocho, Matsuyama,
Ehime 790-8577, Japan}
\email{shidatsu.megumi.wr@ehime-u.ac.jp}

\author[orcid=0000-0001-9911-7038]{Liyi Gu} 
\affiliation{SRON Netherlands Institute for Space Research, Niels Bohrweg 4, 2333 CA Leiden, The Netherlands}
\affiliation{Leiden Observatory, Leiden University, PO Box 9513, 2300 RA Leiden, The Netherlands}
\email{L.Gu@sron.nl}

\author[orcid=0000-0001-9735-4873]{Ehud Behar} 
\affiliation{Department of Physics, Technion, Haifa 32000, Israel}
\affiliation{MIT Kavli Institute for Astrophysics and Space Research, Massachusetts Institute of Technology, Cambridge, MA 02139, USA}
\email{behar@physics.technion.ac.il}

%


%
%

\begin{abstract}

Powerful ionized accretion disk winds are often observed during episodic outbursts in Galactic black hole transients. Among those X-ray absorbers, \fexxvi\ doublet structure (Ly$\alpha_1$+Ly$\alpha_2$ with $\sim 20$eV apart) has a unique potential to better probe the underlying physical nature of the wind; i.e. density and kinematics. We demonstrate, based on a physically-motivated magnetic disk wind scenario of a stratified structure in density and velocity, that the doublet line profile can be effectively utilized as a diagnostics to measure wind density and associated velocity dispersion (due to thermal turbulence and/or dynamical shear motion in winds). Our simulated doublet spectra with post-process radiative transfer calculations indicate that the profile can be (1) broad with a single peak for  higher velocity dispersion ($\gsim 5,000$ km~s$^{-1}$), (2) a standard  shape with 1:2 canonical flux ratio for moderate  dispersion ($\sim 1,000-5,000$ km~s$^{-1}$) or (3) double-peaked with its flux ratio approaching 1:1 for lower  velocity dispersion ($\lsim 1,000$ km~s$^{-1}$)    
in optically-thin regime, allowing various line shape. Such a diversity in doublet profile is  indeed unambiguously  seen in recent observations with XRISM/Resolve at  microcalorimeter resolution. We show that some implications inferred from the model  will help constrain the local wind physics where \fexxvi\ is predominantly produced in a large-scale, stratified wind.



\end{abstract}


\section{Introduction} 

Galactic black hole (BH)  transients\footnote{Low-mass XRBs hosting neutron stars also exhibit X-ray winds. This work, however, does not consider these systems since the physical environment can be quite distinct between the two populations.} are known to undergo episodic outbursts showing a series of distinct transition of accretion state exemplified, most notably, as soft and hard state, characterized by distinct X-ray spectral form \citep[e.g.][]{Fender04, RemillardMcClintock06, Done07, Belloni10, Kylafis12}. During bright soft state when the spectrum is predominantly governed by thermal blackbody emission from an accretion disk,  X-ray winds can often be detected in the form of blueshifted absorption features of multi-ion elements at high ionization state in the spectra. 

Among various absorption lines due to resonance transition, \fexxv\ He$\alpha$ and \fexxvi\ Ly$\alpha$ lines in Fe K complex are the most familiar features in many X-ray observations due  to the high Fe abundances and transition probabilities as well as the natural presence of ionizing X-ray sources responsible for keeping the wind medium highly ionized out to large distances. While the exact  characteristics of these ionized absorbers are quantitatively different from ion to ion, photoionization modeling typically finds that the physical property of the canonical X-ray winds seen in high-inclination BH X-ray binaries (XRBs) are described by hydrogen-equivalent column density of $N_H \sim 10^{21-23}$ cm$^{-2}$, line of sight (LoS) wind velocity of the order of $v \sim 100-1,000$ km~s$^{-1}$ and high ionization state measured by ionization parameter $\xi \equiv L_{\rm ion}/(n r^2) \sim 10^{3-6}$ erg~cm~s$^{-1}$ where $L_{\rm ion}$ is ionizing luminosity in $1-1,000$ Ryd and $n$ is ion number density at distance $r$ \citep[e.g.][]{Miller06,Kallman09,NeilsenHoman12,Trueba19,Miller15,Shidatsu16,Ratheesh21,F21,Parra25}. 

Although a broadband spectral fit of absorption features encompassing a series of different ions (e.g. Ne, O, Mg, Si, S and Fe) is critical for an accurate understanding of a global nature of the disk wind,  Fe K absorbers are often more robustly detected with high significance in many BH XRBs  
with a weak or little low-ionization counterparts in soft X-ray band \citep[e.g.][]{Trueba19} with the exception of GRO~J1655-40 and GRS1915+105 \citep[e.g.][]{Kallman09,Miller15,F21,Ratheesh21,Parra25}. 
%
%
Therefore, H-like and He-like Fe absorbers are often the primary, well-defined atomic features
with high S/N, providing a novel opportunity to better learn the underlying physics of the wind at a level that would otherwise not be obtained with weaker/elusive absorption lines.  


In particular, \fexxvi\  line comes in a doublet structure due to spin-orbit coupling of an electron; 
i.e. $6.952$ keV and $6.973$ keV splitting in a roughly 1:2  ratio of oscillator strengths $f_{\rm ij}$ (between i$^{\rm th}$ and j$^{\rm th}$ electron level)  \citep[e.g.][]{Verner96, Miller15}. The relative strength between Ly$\alpha_1$ line and Ly$\alpha_2$ line is thus generally determined by $f_{\rm ij}$  in optically-thin regime. 
The centroid (trough) line energy is blueshifted from the rest-frame energy intimately reflecting the LoS wind velocity $v$, while its width is primarily reflecting  broadening processes such as (thermal) turbulent motion and/or a dynamical shear motion of the wind \citep[e.g.][]{Tanimoto25}. Therefore, a composite \fexxvi\ doublet structure should in principle be sensitive to a combination of these underlying kinematics. It is hence conceivable that the observed doublet feature (Ly$\alpha_1$ + Ly$\alpha_2$), if sufficiently resolved, may well physically manifest the underlying gas dynamics of the wind responsible for producing the observed \fexxvi\ Ly$\alpha$ absorption  line profile. 

The past analyses in literature have courageously  attempted to identify and resolve this unique feature by maximally utilizing state-of-the-art dispersive data, for example, with third-order {\it Chandra}/HETGS spectra, revealing, for the first time, putative doublets from a number of well-known BH transients \citep[e.g.][]{Miller15}. Although some are relatively evident (e.g. GRO~J1655-40), some are not statistically compelling due to insufficient resolving power and photon statistics  (e.g. GRS~1915+105, H~1743-322 and 4U~1630-472). Hence, the doublet feature is practically treated as a single line even at the first-order HETGS resolution owing to its very narrow separation ($\sim 20$ eV).     
With the successful launch of {\it XRISM} \citep{Tashiro21,Tashiro25} carrying microcalorimeter detectors, {\tt Resolve}, with the energy resolution of 5-7 eV, X-ray astronomy has advanced to the next level where Fe K complex can be further magnified to search for very fine spectral details  including a \fexxvi\ doublet feature.

In this work, we only consider  \fexxvi\ Ly$\alpha$ absorbers in the disk wind of BH XRBs with an exclusive focus on its doublet property that can be  best resolved  with {\it XRISM}/Resolve at statistically significant level. We perform a number of  spectral simulations of the doublet  feature by utilizing a physically-motivated wind model  in an effort to better understand the fundamental wind property; e.g. wind kinematics.  In comparison with the previous studies of \fexxvi\ doublet in literature, our model is more physically driven in the framework of a stratified disk wind rather than a phenomenological  modeling.
In \S 2, we briefly review the essential foundation of our disk  wind model. In \S 3, we present our results and spectral simulations 
relevant for BH transients. We summarize and discuss our findings with implications  in \S 4.

\section{Model Description} \label{sec:model}

\subsection{Essential Nature of the Disk Winds}

The present study is strictly focused on  line profile of outflowing \fexxvi\ absorbers, not broadband absorbers of  multi-ion species. 
It is also becoming apparent that some amount of reprocessing/emission lines are present associated with these transitions, which is not incorporated in this work.
It is thus not quite appropriate to consider and address a mutual interplay among different ions in the current formalism of the wind model employed here.    
In this work, we only make a few necessary assumptions; (1) the  wind is smooth and continuous, (2) launched over a large radial extent of the accretion disk occupying a large solid angle  (this is always valid for a relatively narrow wind shell)  and (3) the property of \fexxvi\  doublet is dependent mainly on the wind density $n$, LoS velocity $v$ and the velocity dispersion (i.e. broadening) $\Delta v$. To this end, we adopt a  stratified wind model with  density gradient  and velocity gradient in steady-state ($\partial / \partial t=0$) under axisymmetry ($\partial/\partial \phi=0$) for simplicity.
Such a configuration is not necessarily restricted to but conformal to certain wind driving mechanisms such as, for example, magnetic driving  \citep[e.g.][]{F10,Kazanas12,Kraemer18,Jacquemin-Ide19, Chakravorty16,Chakravorty23} as considered in this work. 

The essential nature  of this type of  wind  can be captured by two (most) unique characteristics; wind density  $n(r,\theta)$
and LoS outflow velocity $v(r,\theta)$ as a function of LoS distance $r$ and inclination $\theta$ given by  
\begin{eqnarray}
n(r,\theta) \sim n_0 g(\theta)f_{\rm den}  \left(\frac{r}{r_0}\right)^{-p}  ~{\rm and} ~ v(r,\theta) \sim  v_0 h(\theta) f_v \left(\frac{r}{r_0}\right)^{-1/2} \ , \label{eq:eqn1}
\end{eqnarray}
where  $g(\theta)$ and $h(\theta)$ are both angular distributions numerically calculated in the model ($g \sim h \sim 1$ at $\theta \sim \pi/2$ and $\ll 1$ at $\theta \sim 0$; see, e.g., \citealt{F10}). As seen,   
the radial dependence of the wind density is governed by index $p$ 
\citep[e.g.][]{Behar09,Trueba19}\footnote[15]{The index $p$ governs a global radial density distribution of winds, thus our results here are qualitatively insensitive to a specific choice of these numerical value.}, while the LoS velocity obeys Keplerian law. Here, two normalizations have been introduced;  
%
$n_0$ for density at the innermost wind launch radius at $r=r_0$ (conceivably near the innermost stable circular orbit or ISCO) and  $v_0$ for velocity at $r=r_0$. 
Given the  lack of detailed knowledge concerning the innermost disk physics, we have parameterized density by $f_{\rm den}$ and velocity by $f_v$ as a free parameter as shall be addressed later.

Line broadening is conventionally parameterized in the  form of turbulent motion as an independent variable \citep[e.g.][in {\tt xstar} photoionization calculations]{KallmanBautista01}. However, we are motivated in this work to prescribe it as a combination of  thermal and dynamical shear motion,  whichever is locally greater \citep[e.g.][]{Tanimoto25}, such that
\begin{eqnarray}
v_{\rm turb} \equiv f_{\rm turb} \times {\rm max} (v_{\rm sound}, v_{\rm shear}) \ ,
\end{eqnarray}
where $v_{\rm shear}$ is local  shear velocity  determined by velocity gradient over spatial bin $\Delta r$ multiplied by $\Delta r$; i.e. $v_{\rm shear} \equiv (dv/dr)_{\rm mean} \Delta r$
(typically $\sim 5\%-10\%$ of the corresponding proper velocity when $f_v=f_{\rm turb}=1$). In parallel, local sound speed $v_{\rm sound} \simeq 100 \sqrt{T/10^6}$ km~s$^{-1}$ depends on the local plasma temperature $T$ (in units of K) under the ionization equilibrium. To further accommodate the  uncertainty in disk/plasma kinematics,  
we have introduced a parameter $f_{\rm turb}$, which  in fact is a critical factor in this work for  determining the doublet shape as we articulate more below. 


Within the present framework described above, we assume for simplicity $\theta=70\deg$ (relevant for high inclination transients) for a $M=10\Msun$ \sw black hole such that the gravitational radius $r_g$ is approximately $1.5 \times 10^6$ cm and the ISCO radius is  $r =6r_g$, thus $r_{0} \sim 6 r_g$. 
Given that the characteristic shape of \fexxvi\ doublet profile is most sensitive to the degree of velocity dispersion in optically thin regime, we are   motivated to hold other parameters constant. Observations such as absorption measure distribution of X-ray winds often imply $p \simeq 1.0-1.4$ 
\citep[e.g.][]{Behar09,Kallman09,F17,Trueba19,F21}, thus we choose $p=1.2$ here.

Blueshift and line opacity of \fexxvi\ are intimately coupled to wind density and LoS velocity, respectively. We calculate  the line profile by assuming $f_{\rm den}=2$  such that $n_0 f_{\rm den}=2.7 \times 10^{17}$ cm$^{-3}$ \citep[e.g.][]{F17,Trueba19,Ratheesh21,F21}, $n_{\rm in} \equiv n(r_{\rm in},70\deg) \simeq 8.7 \times10^{15}$ cm$^{-3}$ and  $f_v=0.1$ yielding $v_{\rm in} \equiv v(r_{\rm in}, 70\deg) \simeq 0.021c$ (with $c$ being the speed of light) where the subscript ``in" denotes the innermost wind layer that intersects the LoS of $70\deg$ depending on the field geometry.

\begin{deluxetable}{l||cccc|cccc}
\tabletypesize{\small} \tablecaption{Characteristic Quantities in the Wind Model} \tablewidth{0pt}
\tablehead{Variable & $r_{\rm in}$ [cm] $^\dagger$ & $n_{\rm in}$ [cm$^{-3}$]  & $v_{\rm in}/c$ $^\sharp$ & $v_{\rm turb,in}/c$ $^\flat$  & $v_r/v_K(6r_g,90\deg)$ $^\ddagger$ & $v_\phi/v_K(6r_g,90\deg)$ $^\ddagger$ & $v_z/v_K(6r_g,90\deg)$ $^\ddagger$    
}  
\startdata
Value &   $1.5 \times 10^7$  &  $8.7 \times 10^{15}$  & $0.021$ & $0.011$ & 0.07 & 0.92 & 0.06 & \\  \hline 
%
%
\enddata
\label{tab:tab1}
\vspace*{0.2cm}
$^\dagger$At the  innermost wind layer ($r=r_{\rm in}$ and $\theta=70\deg$) that intersects $70\deg$ LoS. 
$^\ddagger$At the innermost launch site ($r=6r_g$ and $\theta=90\deg$) in the equator ($90\deg$). $c$ is the speed of light and $v_K$ is the Keplerian velocity.  
$^\sharp$ $f_v=0.1$. $^\flat$ $f_{\rm turb}=1$.
\end{deluxetable}
%

\begin{figure}[t]
\begin{center}
\includegraphics[trim=0in -0in 0in
0in,keepaspectratio=false,width=2.9in,angle=-0,clip=false]{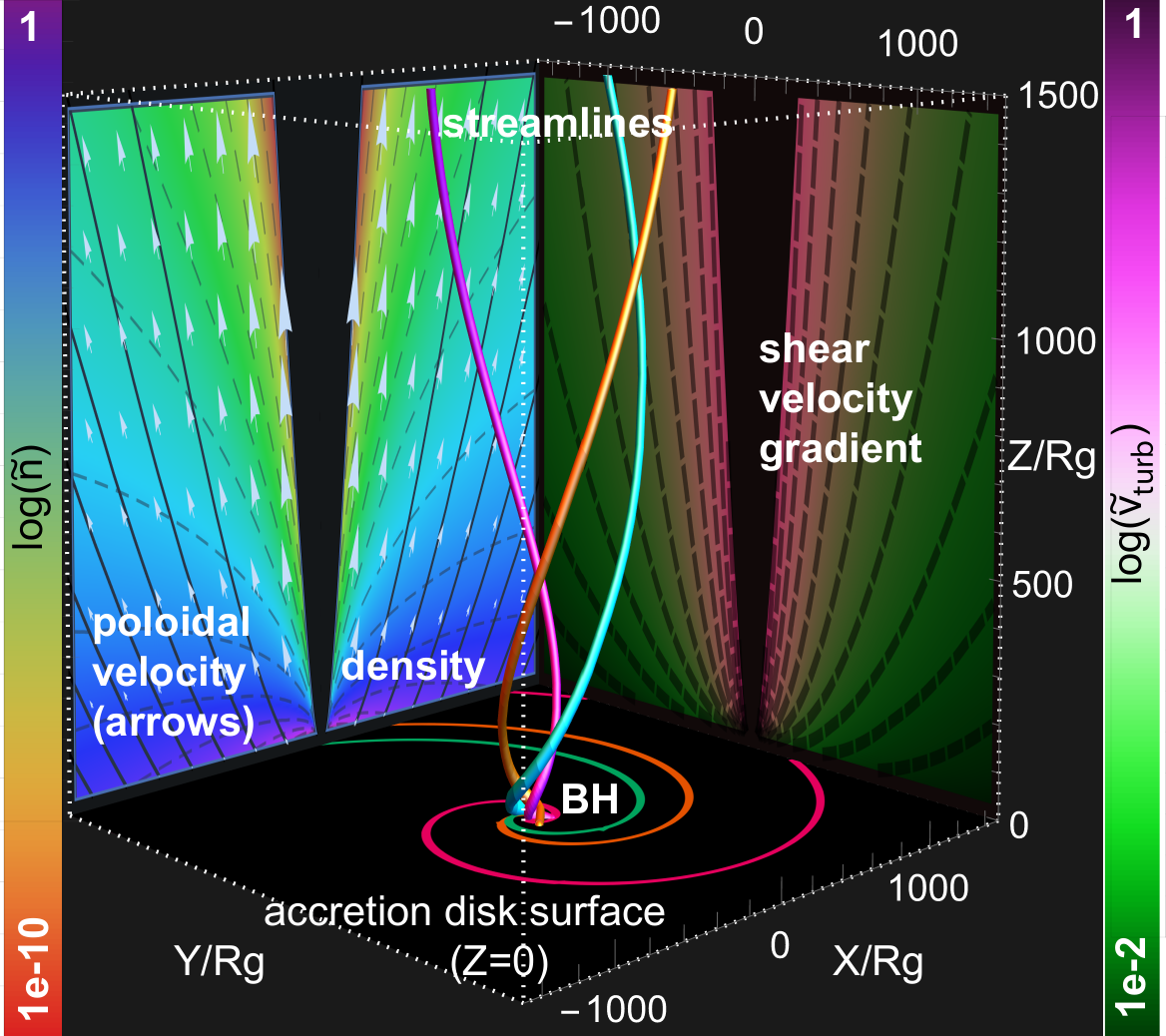}\includegraphics[trim=0in -0in 0in
0in,keepaspectratio=false,width=3.4in,angle=-0,clip=false]{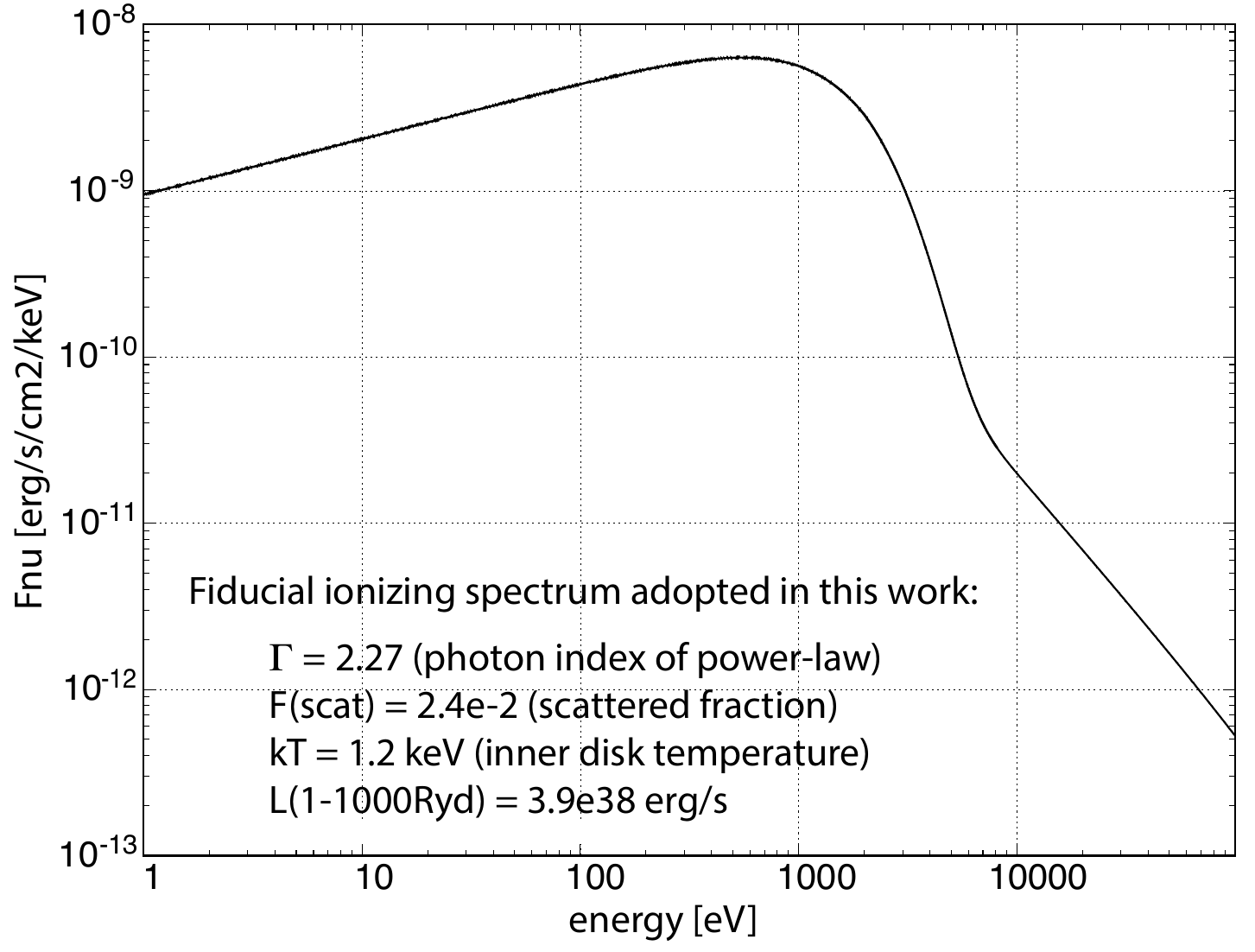}
\end{center}
\caption{(a) Three-dimensional rendering of three individual streamlines launched magnetically from different locations. On X-Z plane, we show the poloidal distribution of the normalized wind density (color-coded) $\log \tilde{n}(r,\theta)$,  density contours (dashed) and the poloidal projection of magnetic field lines (thick solid). Y-Z plane shows the distribution of the normalized poloidal turbulent velocity  $\log \tilde{v}_{\rm turb}(r,\theta)$ (color-coded)  with its contours (dashed). Toroidal projection of the streamlines is shown on X-Y plane. (b) Intrinsic X-ray ionizing spectrum (relevant to bright soft state of BH XRBs dominated by thermal radiation from the disk) adopted for post-process radiative transfer calculations with {\tt xstar} in this work \citep[see,][for details]{ShidatsuDone19,Tanimoto25}. }
\label{fig:f1}
\end{figure}

With the choice of these parameters, we show a fiducial magnetic wind solution in {\bf Figure~\ref{fig:f1}a} where three-dimensional rendering of three individual streamlines is demonstrated. On X-Z plane, we show the poloidal distribution of the normalized wind density (color-coded) $\log \tilde{n}(r, \theta)$ where  $\log \tilde{n}(r,\theta) \equiv \log (n/n_{\rm min})/\log (n_{\rm max}/n_{\rm min})$, density contours (dashed),  the poloidal projection of magnetic field lines (thick solid), and the poloidal wind velocity (white arrows). Y-Z plane shows the distribution of the normalized  turbulent velocity gradient $\log \tilde{v}_{\rm turb}$ (color-coded) with its contours (dashed) where $\log \tilde{v}_{\rm turb}(r,\theta) \equiv \log (v_{\rm turb}/v_{\rm turb,min})/\log (v_{\rm turb,max}/v_{\rm turb,min})$.
Toroidal projection of the streamlines is shown on X-Y plane (at $Z = 0$). 
With $f_{\rm turb}=1$,  we find $v_{\rm turb}(r_{\rm in},70\deg) \sim \Delta v(r_{\rm in},70\deg) \simeq 0.011c \simeq 3,278$ km~s$^{-1}$, which is $\sim 50\%$ of the wind velocity $v_{\rm in}  \simeq 0.021c$ with $f_v=0.1$. 
The specific values  of the wind property is highly model-dependent. However, because the local turbulent velocity, $v_{\rm turb}$, is uniquely given by a stratified velocity field in equation~(\ref{eq:eqn1}), we numerically obtain the following values of dispersion (broadening) ranging broadly from  $f_{\rm turb} = 0.1$ ($v_{\rm turb, in} \sim 328$ km~s$^{-1}$) to $4 ~(v_{\rm turb,in} \sim 13,112$ km~s$^{-1}$) 
%
%
for our spectral calculations. 
Note that $v_{\rm turb,in}$ represents local turbulent velocity (i.e. dispersion) at $r=r_{\rm in}$, not where \fexxvi\ column is peaked (see the definition of $r_{\rm peak}$ later in \S 2.2).   
%
%
The characteristic wind quantities considered in this study are listed in {\bf Table~\ref{tab:tab1}}.



\subsection{Post-Process Photoionization Balance}

Numerically obtained wind solutions are  illuminated by ionizing photons. In this work, we are interested in bright soft state where X-ray winds are often most pronounced in the observed spectra dominated by the thermal disk emission of BH XRBs. For post-process photoionization calculations with {\tt xstar} \citep[][]{KallmanBautista01}, we provide a fiducial (de-absorbed) ionizing spectrum (see {\bf Fig.~\ref{fig:f1}b}) to qualitatively mimic the characteristic soft state spectrum of typical BH transients following the work by \cite{ShidatsuDone19} and \cite{Tanimoto25} in their wind modeling. 

\begin{figure}[t]
\begin{center}
\includegraphics[trim=0in -0in 0in
0in,keepaspectratio=false,width=3.8in,angle=-0,clip=false]{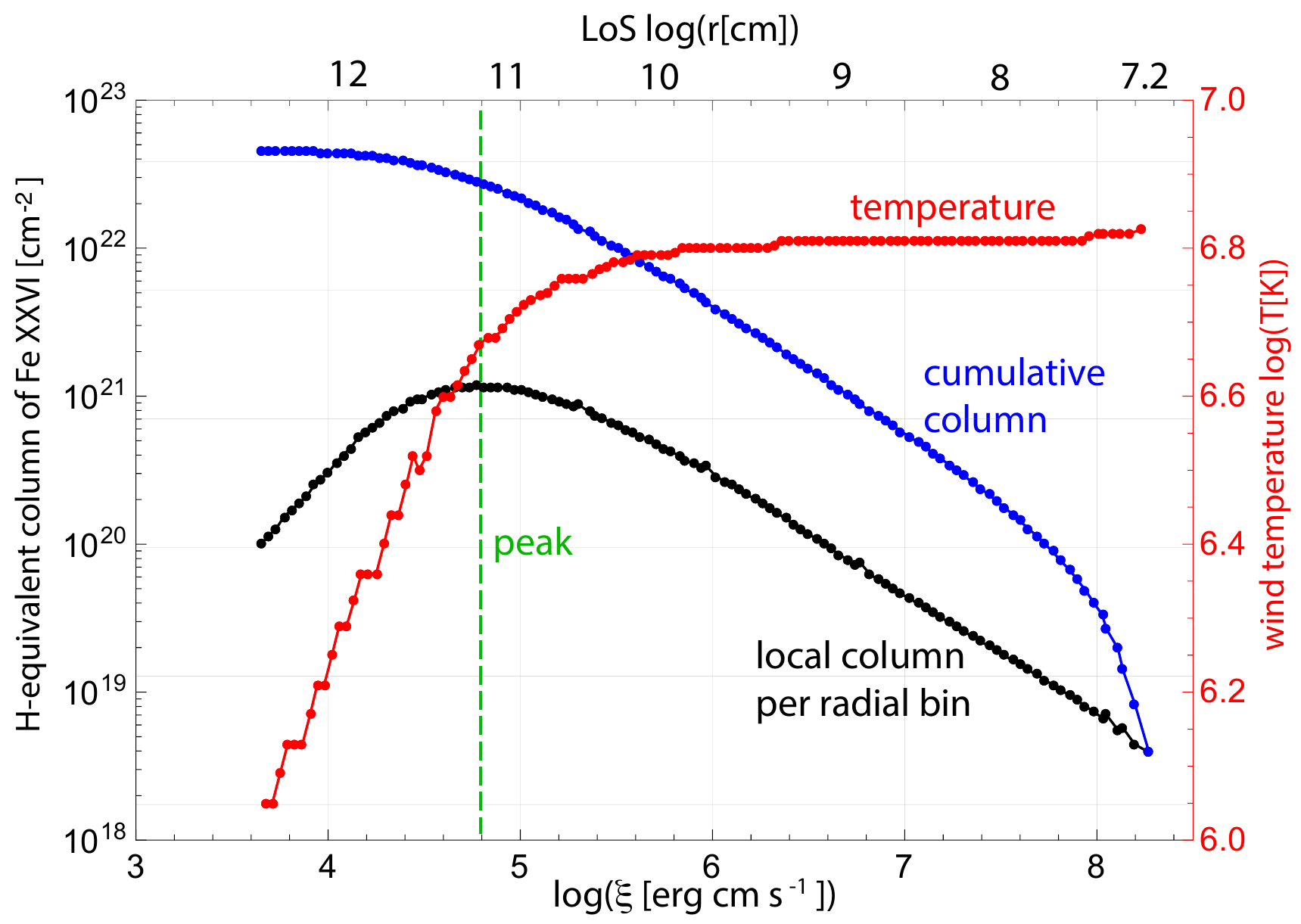}\includegraphics[trim=0in -0in 0in
0in,keepaspectratio=false,width=3.5in,angle=-0,clip=false]{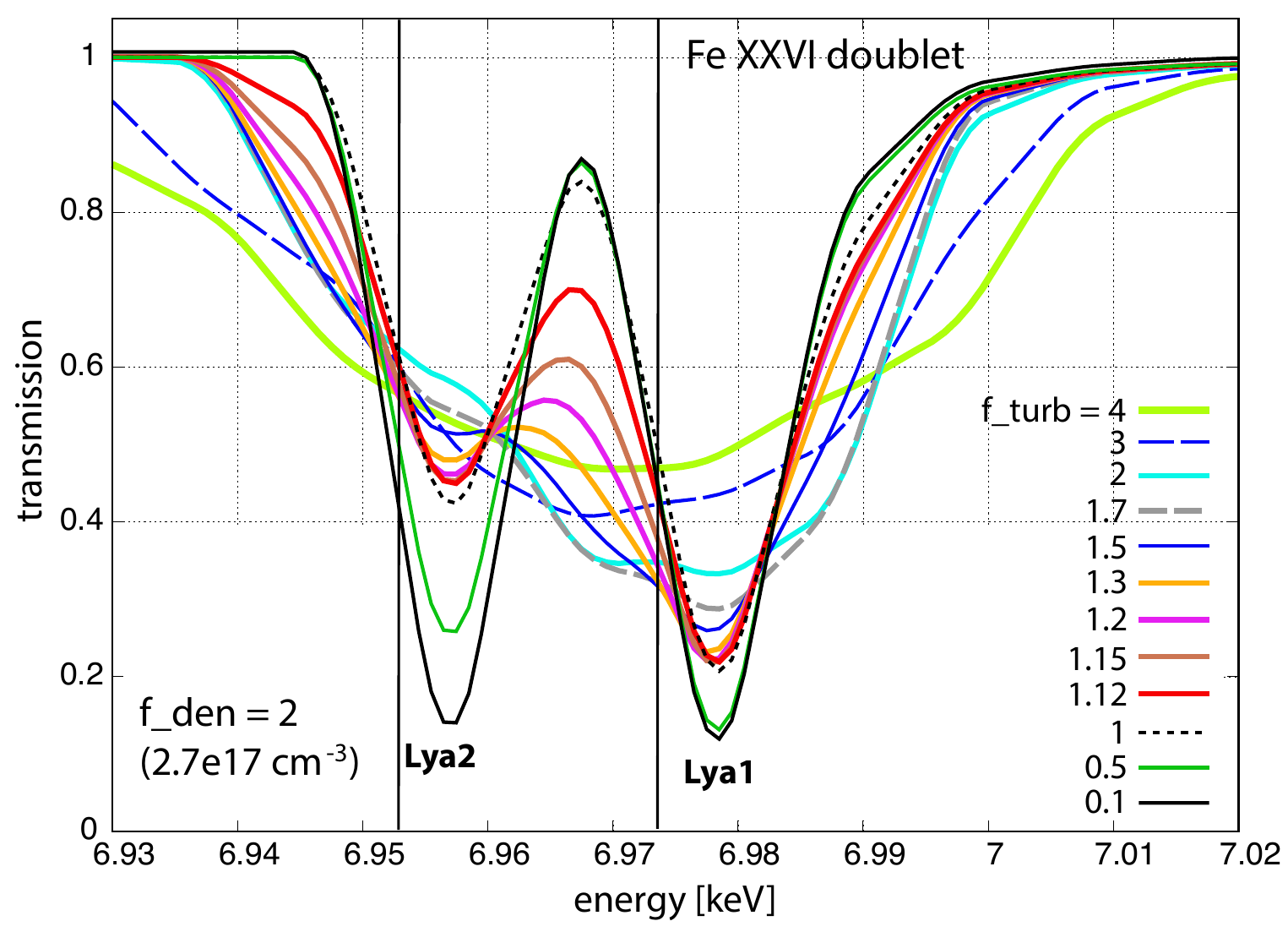}
\end{center}
\caption{(a) Distribution of \fexxvi\ (hydrogen-equivalent) local column $N_H$ per radial bin (dark), cumulative column (blue)  and the wind temperature $T$ (red) as a function of ionization parameter $\xi$  (and LoS distance $\log r[\rm cm]$) for $\theta=70\deg$ (LoS). Vertical dashed line denotes the peak (at $r=r_{\rm peak}$) point where \fexxvi\ column is maximum. (b) Calculated doublet structure of \fexxvi\ absorption line corresponding to {\bf Figure~\ref{fig:f2}a} for various broadening dispersion factor $f_{\rm turb}$. Rest-frame Ly$\alpha_1$ and Ly$\alpha_2$ energies are denoted by vertical lines. We assume  $f_{\rm den}=2$ (i.e. $n_o f_{\rm den}=2.7 \times 10^{17}$ cm$^{-3}$), $p=1.2$ and  $f_v=0.1$ under optically-thin regime. 
}
\label{fig:f2}
\end{figure}

As a result of radiative transfer calculations assuming $L_{\rm ion} = 3.9 \times 10^{38}$ erg~s$^{-1}$ (i.e. $\sim 30\%$ of Eddington rate)  in $1-10^3$ Ryd, we obtain a distribution of ionic columns for various elements along the LoS (i.e. $\theta = 70\deg$ here). 
Among those, we show in {\bf Figure~\ref{fig:f2}a} the LoS distribution of local \fexxvi\ (hydrogen-equivalent) column $N_H$ (dark)  over a spatial grid in our computation, cumulative column (blue) and the wind temperature $T$ (red) as a function of ionization parameter $\xi$  for  $f_{\rm den}=2$ (such that $n_o f_{\rm den}=2.7 \times 10^{17}$ cm$^{-3}$) with $p=1.2$ and $f_v=0.1$.  The corresponding LoS distance $\log r[\rm cm]$ is also shown for reference.  As the gas temperature gradually decreases with decreasing $\xi$, the column production is seen to monotonically increases peaking at an optimal ionization level.
The wind velocity decreases accordingly with distance as X-ray propagates outward intercepting different parts of the global outflows \citep[see {\bf Fig.~\ref{fig:f1}a}; also][for details]{F10}. 

We find the following characteristic values where  local \fexxvi\ column  becomes maximum at the peak point (vertical dashed line in {\bf Fig.~\ref{fig:f2}a}); i.e. $N_H^{\rm peak} \simeq 1.2 \times 10^{21}$ cm$^{-2}$ at LoS distance of $r_{\rm peak} \simeq 1.8 \times 10^{11}$ cm $\simeq 1.2 \times 10^5 r_g$ when $\log \xi_{\rm peak} \simeq 4.8, v_{\rm peak}  \simeq 57$ km~s$^{-1}$, $T_{\rm peak} \simeq 10^{6.6}$ K. 
It is reminded that velocity broadening by turbulence $v_{\rm turb}$ also responds to the change in the gas temperature (for thermal turbulence via $v_{\rm sound}$) as well as dynamical velocity gradient (for shear motion via $v_{\rm shear}$) as described earlier.  
As an example, the wind model yields $v_{\rm shear}^{\rm peak} \simeq 30$ km~s$^{-1}$ and $v_{\rm sound}^{\rm peak} \simeq 214$ km~s$^{-1}$ with $f_{\rm turb}=1$.  
Although  the cumulative column amounts to the order of $\lsim 10^{23}$ cm$^{-2}$ with increasing distance, the local wind  remains optically-thin, and the latter matters most to local ionization balance.

\section{Results} \label{sec:results}

With the calculated \fexxvi\ column density in conjunction with the simulated disk wind of a stratified density and velocity structure described in \S 2 within the framework of MHD-driving, we calculate the transmission $e^{-\tau(r,E)}$ where $\tau(r,E)$ is line optical depth (as a function of distance $r$ and photon energy $E$) defined as $\tau \equiv N_{\rm ion} \sigma_{\rm abs}$; i.e. $N_{\rm ion}$ is the ionic column density of \fexxvi\ and $\sigma_{\rm abs}$ is the photo-absorption cross section of an ionized ion. It is given by 
%
$\sigma_{\rm abs} = 0.01495 (f_{\rm ij}/v_{\rm turb}) H(a,u)$
%
assuming that the  line profile is well approximated locally by the Voigt function $H(a,u)$ where $a$ is related to Einstein coefficient $A_{\rm ij}$ and $v_{\rm turb}$, while a variable $u$ also depends on $v_{\rm turb}$  \citep[e.g.][]{Mihalas78, Kotani00, Hanke09}. 
Considering a situation where the wind medium is locally optically-thin  (i.e. $N_H \lsim 10^{21}$ cm$^{-2}$),  canonical doublet line ratio Ly$\alpha_1$/Ly$\alpha_2$ is dictated  uniquely by  $f_{\rm ij}$ while its intensity is dominated by  wind density. However, 
the effect of {\it kinematics} in tandem with a physically-motivated  wind model is not well studied to date beyond a phenomenological  description. This is the primary theme of this work. 

\begin{figure}[t]
\begin{center}
\includegraphics[trim=0in -0in 0in
0in,keepaspectratio=false,width=3.5in,angle=-0,clip=false]{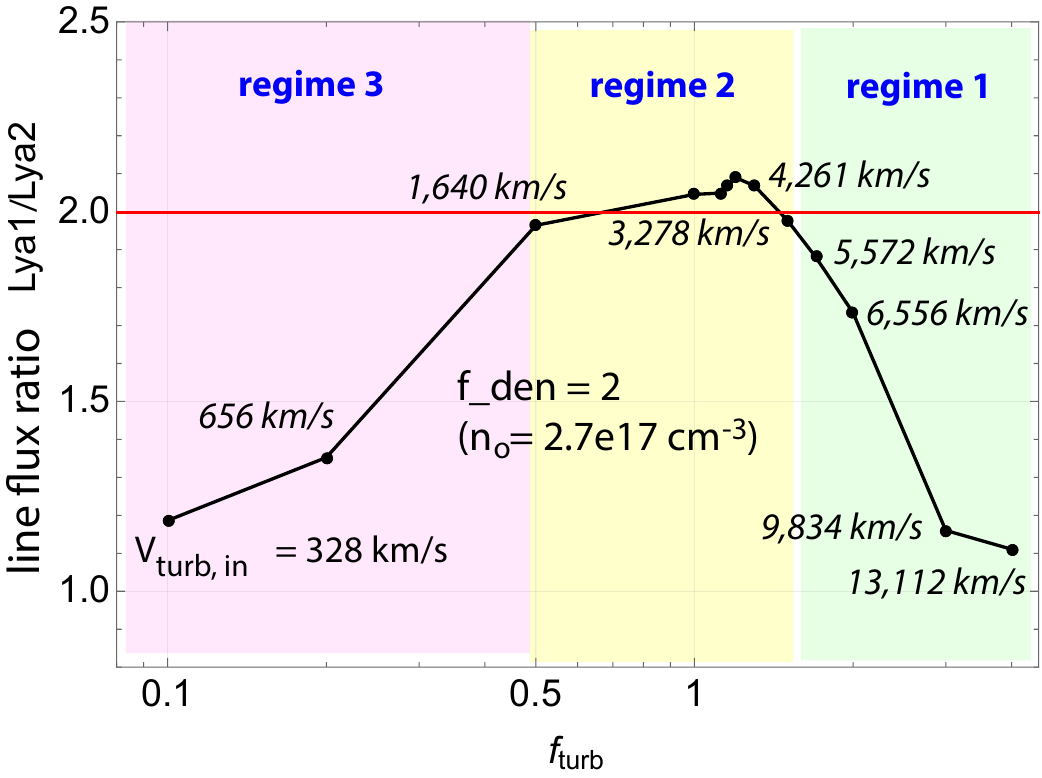}\includegraphics[trim=0in -0in 0in
0in,keepaspectratio=false,width=3.6in,angle=-0,clip=false]{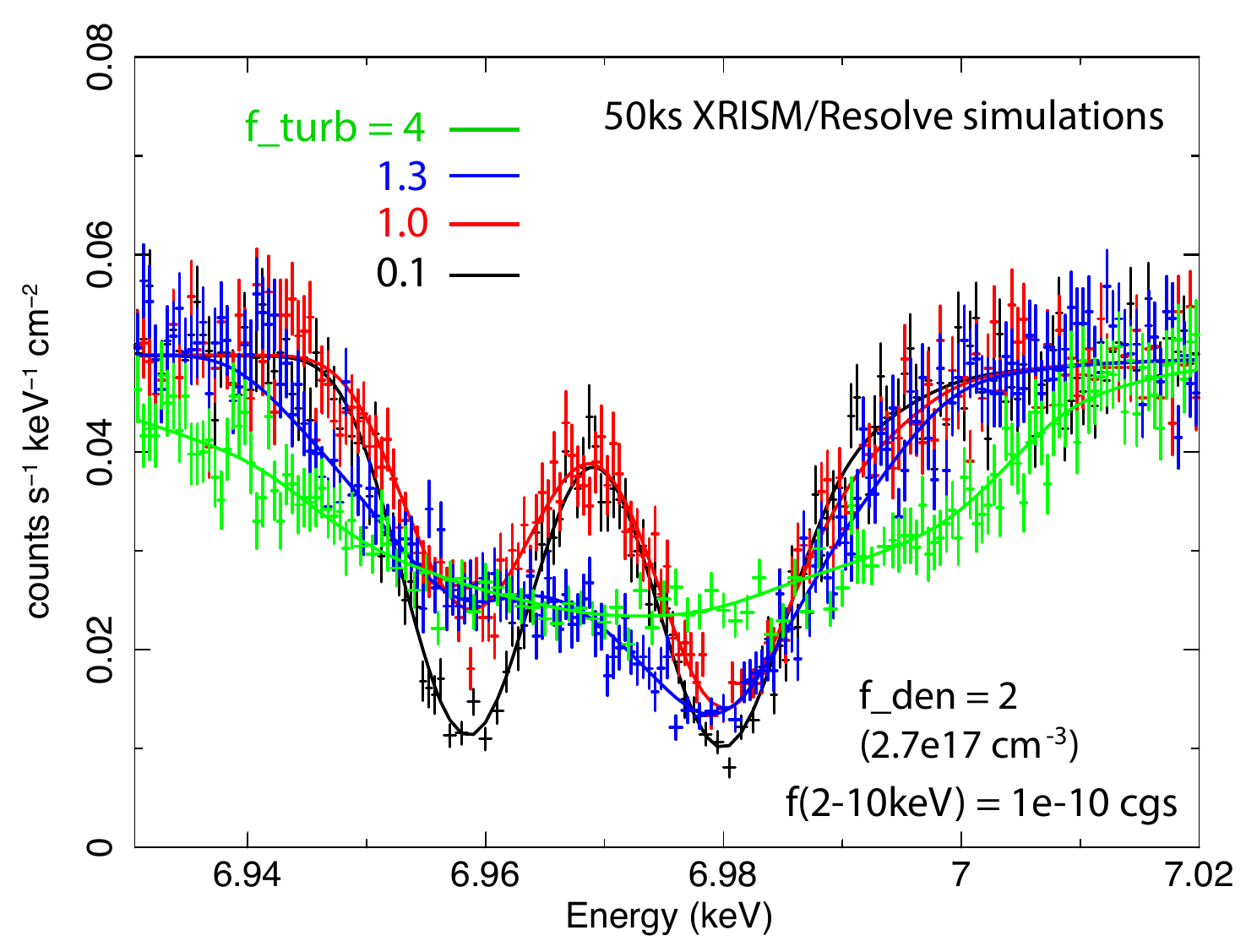}
\end{center}
\caption{(a) Calculated line ratio (Ly$\alpha_1$/Ly$\alpha_2$) corresponding to {\bf Figure~\ref{fig:f2}b} as a function of broadening dispersion $f_{\rm turb}$ with $f_v=0.1$ and $f_{\rm den}=2$ where red line denotes the expected 1:2 ratio under optically-thin regime. Denoted is the corresponding $v_{\rm turb,in}$ value. (b) Simulated 50ks XRISM/{\tt Resolve} spectrum with $f_{\rm den}=2$ and $f_v=0.1$ assuming 2-10 keV flux of $10^{-10}$ [cgs] for various dispersion; $f_{\rm turb} = 0.1$ (dark), $1$ (red), $1.3$ (blue) and $4$ (green). 
}
\label{fig:f3}
\end{figure}

In {\bf Figure~\ref{fig:f2}b} we show the predicted \fexxvi\  doublet spectrum for various $f_{\rm turb}$ assuming $f_{\rm den}=2$ (such that $n_o f_{\rm den}=2.7 \times 10^{17}$ cm$^{-3}$) for $p=1.2$ and $f_v=0.1$ with the understanding that a major driving factor to shape the doublet profile is $f_{\rm turb}$ under optically-thin regime (i.e. $f_{\rm den} = 2$ yields  $N_H^{\rm peak} \lsim 10^{21}$ cm$^{-2}$; see {\bf Fig.~\ref{fig:f2}a}). 
We thus focus in the present work exclusively on the the role of $f_{\rm turb}$ for doublet line profile. To complement this result, {\bf Figure~\ref{fig:f3}a} shows the characteristic line flux ratio, Ly$\alpha_1$/Ly$\alpha_2$,  measured at the trough energy of each line as a function of $f_{\rm turb}$ in the model.   


It is clear that a diverse doublet structure can be attained by different broadening factor even though the the rest of the intrinsic property of the wind (e.g. density, ionization structure and LoS wind velocity) still remains the same. 
With large $f_{\rm turb}$ (e.g. $f_{\rm turb} \gsim 1.5$ or $v_{\rm turb,in} \gsim 4,917$ km~s$^{-1}$), the doublet feature is practically blended, exhibiting an asymmetric broad line shape with the trough energy more weighted by Ly$\alpha_1$ line. Line flux ratio, Ly$\alpha_1$/Ly$\alpha_2$, while calculated to be well below 2, thus would not mean much physically in this case. 
With decreasing  broadening factor, the extreme line blending is somewhat relaxed, resulting in the presence of relatively noticeable Ly$\alpha_1$ trough, although the doublet line ratio is not quite 1:2. This is {\it regime 1}.

For intermediate factor (e.g. $f_{\rm turb} \sim 0.5-1.5$ or $v_{\rm turb,in} \sim 1,640 - 4,917$ km~s$^{-1}$), both lines continue to grow to be independently noticed with little blending. Note that the transmission between Ly$\alpha_1$ and Ly$\alpha_2$  (around $E \sim 6.965$ keV) increases  with decreasing $f_{\rm turb}$ as a  consequence of narrower line width. Both lines are formed deeper such that the line ratio is stable around 1:2, as expected, independent of velocity dispersion $f_{\rm turb}$. This is {\it regime 2}. Note a slight discrepancy from 1:2 ratio due to numerical artifact in our spectral calculations.

When velocity broadening  becomes even smaller (e.g. $f_{\rm turb} \lsim 0.5$ or $v_{\rm turb,in} \lsim 1,640$ km~s$^{-1}$), the trough of Ly$\alpha_2$ line is further deepened approaching the same level as that of Ly$\alpha_1$. This is  caused primarily by the significant change of the cross section $\sigma_{\rm abs}$  that is  proportional to $1/\Delta v \propto f_{\rm turb}^{-1}$ as addressed above, despite the fact that the gas density remains unchanged. As a result, the line ratio deviates from 1:2. This is {\it regime 3}. 

Be mindful, however, that the calculated shear velocities $v_{\rm turb,in}$ at $r=r_{\rm in}$ (i.e. highest shear along the LoS) in the above may vary depending on the exact wind density ($f_{\rm den}$) and blueshift ($f_v$), although a general trend presented here (i.e. regimes 1 - 3) is qualitatively robust and persistent. 

We demonstrate observational feasibility in {\bf Figure~\ref{fig:f3}b} by simulating 50ks XRISM/{\tt Resolve} spectra for various $f_{\rm turb}$ corresponding to theoretical calculations in {\bf Figure~\ref{fig:f2}b}; i.e. $f_{\rm turb} = 0.1$ (dark), $1$ (red), $1.3$ (blue) and $4$ (green). 
We assume here a conservatively low  flux of $f_{\rm 2-10keV} = 10^{-10}$ erg~cm$^{-2}$~s$^{-1}$ in $2-10$ keV \citep[e.g.][]{Kallman09,Miller15,ShidatsuDone19,F21}.  It is clearly seen that the characteristic doublet structure is robustly distinguished among different cases at microcalorimeter resolution.

\begin{figure}[t]
\begin{center}
\includegraphics[trim=0in -0in 0in
0in,keepaspectratio=false,width=3.5in,angle=-0,clip=false]{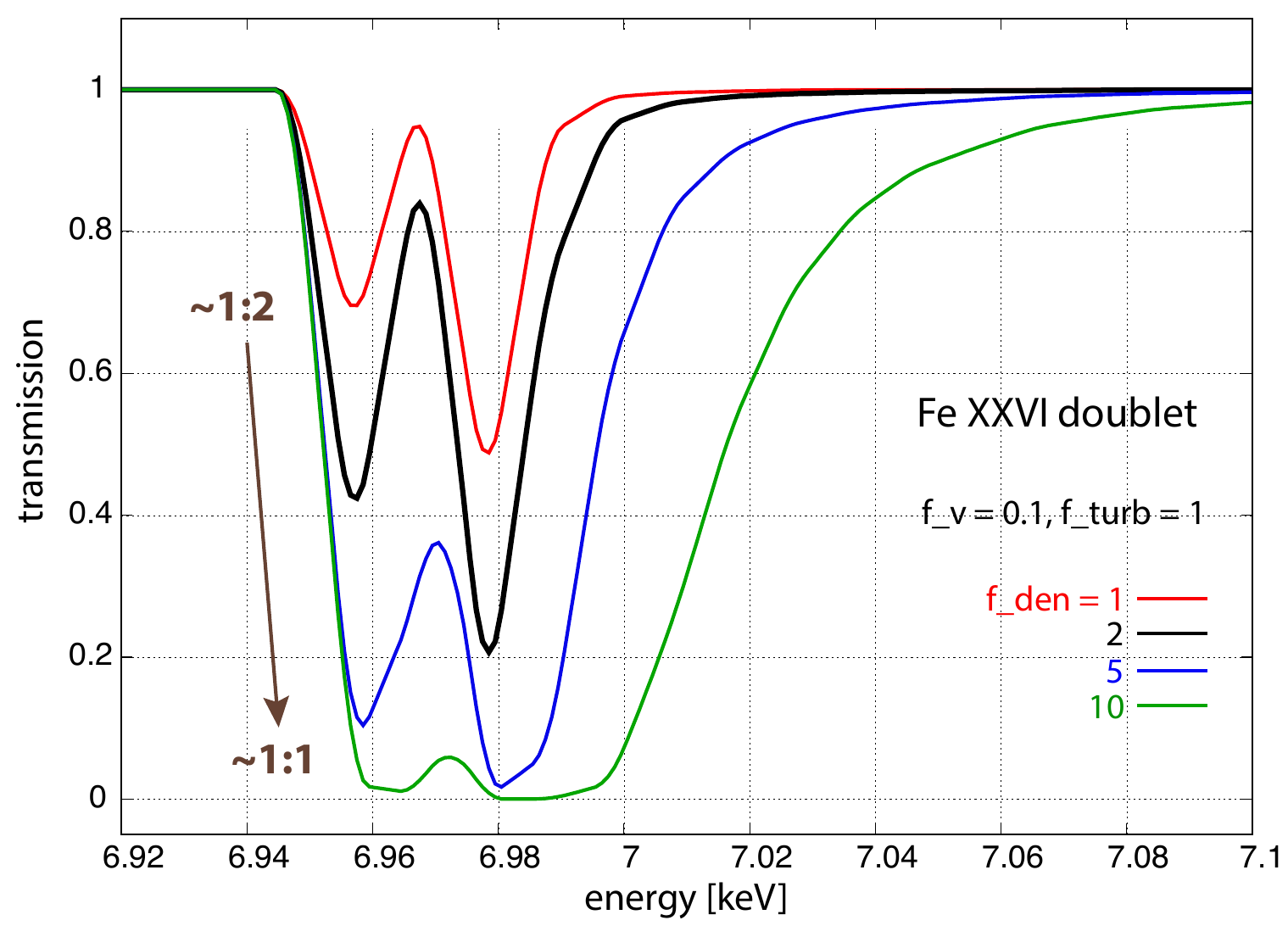}
\end{center}
\caption{Similar to {\bf Figure~\ref{fig:f2}b} but for various $f_{\rm den}$ ranging from optically-thin regime (e.g. $f_{\rm den} \lsim 2$ or $N^{\rm peak}_H \lsim 10^{21}$ cm$^{-2}$ yielding roughly 1:2 ratio) to optically-thick regime (e.g. $f_{\rm den}=10$ or $N^{\rm peak}_H \gsim 10^{22}$ cm$^{-2}$ leading approximately to 1:1) in the same wind  model with $f_v=0.1$ and $f_{\rm turb}=1$. Note that the cumulative column exceeds $\sim 10^{23}$ cm$^{-2}$ for $f_{\rm den} \gsim 5$.
}
\label{fig:thick}
\end{figure}

To briefly explore an optically-thick case, on the other hand, we increase the wind density (with $f_{\rm den}$) with everything else being fixed for radiative transfer calculations ($f_v=0.1$ and $f_{\rm turb}=1$) and compute doublet structure in {\bf Figure~\ref{fig:thick}}. Note that the cumulative column exceeds $\sim 10^{23}$ cm$^{-2}$ for $f_{\rm den} \gsim 5$ as opposed to those shown in {\bf Figure~\ref{fig:f2}}. With increasing density, the feature is clearly seen to be deepened with  line ratio varying from 1:2 to 1:1 as the feature gets almost saturated with higher density, being in a broad agreement with the previous literature  \citep[e.g.][]{Tsujimoto25}. Note that the overall doublet feature in our calculations becomes slightly more blueshifted with increasing density $f_{\rm den}$ because the  ionization front  to produce \fexxvi\ Ly$\alpha$ ion shifts radially inwards along LoS distance under the ionization equilibrium  due to $\xi \propto 1/(n r^2) \propto 1/(f_{\rm den} r^{2-p})$ where $p=1.2(<2)$. Therefore, the ionizing front compensates for the change by $f_{\rm den}$.

In summary, our calculations show that the doublet shape can qualitatively reflect 
the underlying velocity dispersion  in optically-thin regime ({\bf Figs.~\ref{fig:f2}-\ref{fig:f3}}), while density would additionally come into play to affect the profile  in optically-thick regime ({\bf Fig.~\ref{fig:thick}}) which is beyond the scope of the current work. A more detailed calculation will be deferred to a future work.  It should be reminded again that the quantitative estimates of $v_{\rm turb,in}$ above depend on specific wind solutions. Nonetheless, the qualitative trend  is unique and generic to the present model. This points to the fact that treatment of the turbulence significantly affects the shape of the doublet, and should be considered in combination with the other physical parameters we know of. 


\section{Summary \& Discussion} \label{sec:summary}

In this work, we perform radiative transfer calculations with a type of disk winds characterized by stratified density and velocity distributions such as those with  magnetic-driving  in the context of BH XRBs. For spectral modeling, we particularly consider  bright soft accretion state  where powerful disk winds are occasionally witnessed in the form of blueshifted absorption lines of highly ionized ions like \fexxvi. 
Our approach to model \fexxvi\ doublet feature is thus physically motivated based on a well-defined disk wind theory rather than a phenomenological treatment of mutually-decoupled ions  with Gaussian/Voigt functions,  which can be naturally extended to simulate AGN winds as well with proper calibration of ionizing spectrum and density normalization.

In this work  where we are focused on BH XRB winds, it is  demonstrated by spectral simulations that  \fexxvi\ Ly$\alpha$ doublet structure may in principle reflect the underlying wind kinematics, namely, turbulent velocity (or dispersion) for broadening effect.  In the current framework, we show that the degree of turbulence (a combination of thermal turbulence $v_{\rm sound}$ and dynamical shear $v_{\rm shear}$) can imprint a unique doublet profile with a diverse spectral appearance at low optical depth; A broad single line with no distinguishable troughs can be formed if turbulence is sufficiently large ({\it regime 1}). In contrast, the feature can be a well-defined doublet with the expected 1:2 flux ratio if the turbulence is in a moderate range ({\it regime 2}). Under the wind with a  small turbulence, Ly$\alpha_2$ trough can be enhanced to the extent that Ly$\alpha_1$ and Ly$\alpha_2$ are almost comparable to each other in transmission  ({\it regime 3}). Hence, we argue, from the standpoint of the present simplistic wind model, that a diversity of  \fexxvi\ doublet feature may be a tangible manifestation of the underlying 
wind turbulent properties, 
which would otherwise be hard to access observationally.

Once again, it should be reemphasized that the predicted doublet structure of Ly$\alpha$ \fexxvi\  is mostly reflecting a local  condition of the optically-thin disk wind (i.e. thermal condition, density and local kinematics) where \fexxvi\ column is predominantly produced. Therefore, it is not appropriate for the current model to make a prediction of a global aspect of the disk wind such as a  connection to the property of low-ionization X-ray absorbers in soft band (e.g. Ne, Mg, Si and S) or UV counterparts. The implications inferred from the present work hence do not  extend beyond Fe K complex.

In our calculations, we hold constant a set of the parameterized variables at reasonable values; i.e. $p=1.2$ (from the observed absorption measure distribution in typical X-ray winds; e.g. \citealt{Behar09,F17,Trueba19,F21}), $f_v=0.1$ and $f_{\rm den}=2$ corresponding to $n_0 f_{\rm den} \simeq 2.7  \times 10^{17}$ cm$^{-3}$ (as implied from a sample of canonical X-ray winds in BH XRBs; e.g. \citealt{Miller15,F17,Ratheesh21,F21}). 
%
A choice of different values of $p$ and $\theta$ would yield a different line depth and the overall blueshift, but no significant change is qualitatively made in the variation of doublet profile in optically-thin winds. 


%
%
%

Specifically, recent {\it XRISM}/Resolve observations have revolutionized  our understanding of the true spectral appearance of \fexxvi\ doublet profile at microcalorimeter resolution. During {\it XRISM} performance verification (PV) phase, a number of sources have been observed to unambiguously reveal a detailed line profile (e.g. {\it XRISM} collaboration, 2025, in private communication); e.g. a single-peaked, asymmetric broad profile (e.g. 4U~1630-472; \citealt{Miller25}), a double-peaked, asymmetric  profile (e.g. MCG-6-30-15, NGC~4151, GX~13+1  and 4U~1624-490), and a nearly symmetric, deep double-trough profile (e.g. 4U~1916-053), although it can also be (highly) variable over some timescale \citep[see also][]{Tsujimoto25}. Owing to state-of-the-art energy resolution  of {\tt Resolve} instruments ($\sim 5-7$ eV $\ll 20$ eV of doublet separation), it is now realistically feasible to robustly constrain the doublet shape with high statistical significance, which was not quite achievable at the same level even with dispersive instruments like  {\it Chandra}/HETGS and {\it XMM-Newton}/RGS \citep[e.g.][]{Miller15}. 
Although part of these disk winds can be nearly optically-thick, one can attempt to derive a rough estimate on values of wind density and velocity dispersion expected within the current framework. For example, our model suggests that the observed doublet line profiles of \fexxvi\ from the above BH XRBs  seem to be reasonably accounted for by high column of $N_H > 10^{22}$ cm$^{-2}$ and  low or moderate velocity dispersion of $f_{\rm turb} \sim 0.1-1.3$ (equivalent to hundreds to thousands of km~s$^{-1}$) based on the published XRISM/Resolve data (Tsujimoto et al. 2025).
In addition, exploratory studies of Fe K complex of other galactic sources have been recently conducted with XRISM/Resolve data  suggesting high  column density (on the order of $N_H \sim 10^{23}$ cm$^{-2}$) and relatively low/moderate turbulent velocity (on the order of $100-1,000$ km~s$^{-1}$) of highly ionized gas in MAXI~J1744-294\footnote[16]{Specifically for MAXI~J1744-294, we are aware of an ongoing  analysis of the same XRISM/Resolve data  conducted by an independent group  (soon to be submitted) that may lead to a different interpretation and conclusion, reflecting a complex state of the observed spectral components of this source.} \citep{Chatterjee25} and  GX~340+0
\citep{Chakraborty25}, for example, in a broad agreement with our model estimates.  Caution should be taken, however, because scattered emission may play a noticeable role in reshaping the line profile for these sources, which is not considered in the current model. 
For such an optically thick material, a more robust spectral modeling of doublet feature with XRISM/Resolve will be a future work built upon the current model.

In comparison with our present work, 
%
\cite{Tsujimoto25}, for example, have presented a detailed spectral analysis of ionized outflows in Circinus X-1 with XRISM data based on radiative transfer calculations using {\it cloudy}, showing a transition of \fexxvi\ Ly$\alpha$ doublet line ratio from optically-thin to optically-thick regime.  Our calculations are broadly consistent with their findings in terms of 1:2 line ratio, and we further suggest that even the canonically stable 1:2 ratio in optically-thin regime can be substantially altered depending on the wind kinematics of turbulence. 
Correlations between \fexxvi\ feature and the spectral evolution of BH XRBs have also been investigated in detail \citep[e.g.][for 4U~1630-47]{Parra25}, which is another interesting proxy to be addressed in our future work.

While line intensity and line ratios are affected by self-absorption, other external factors (e.g. absorption by a foreground component, saturation, covering factor, and scattered emission) would make it difficult to fully disentangle with just doublet feature, thus further requiring multi-ion lines. On the other hand, the line center remains a relatively isolated and robust indicator of the bulk velocity. The line width, before saturation effects become dominant, may be used to constrain the integrated micro-motion (i.e. turbulence) along LoS. Nonetheless, in a situation like {\it regime 1} where lines become very broad, an accurate constraint on even the line center becomes challenging. Therefore, under conditions such as {\it regimes 2-3} where the lines are still narrow and unsaturated, both the bulk velocity and  turbulence can be best measured.

Our study here is partially restricted to a situation where LoS optically-thin plasma singly plays a major role in shaping doublet feature. However, it is plausible that additional effects could independently affect the line profile. For example, part of the ionized wind medium may well be in near/super-Compton thick state such that scattering becomes important. Such a high density region can be found, for instance, at the base of disk winds especially in high inclination BH XRBs including dipping/eclipsing sources (e.g. GRO~J1655-40 and GRS~1915-105; see \citealt{Miller20}). In general, scattering process also becomes inevitable when X-ray outflows get  externally obscured by some materials (e.g. dense obscurers or failed winds). This condition may be reminiscent of Compton-thick Seyfert AGNs \citep[e.g.][]{Marchesi22}. In that case, scattered emission component becomes non-negligible,  making a substantial contribution to a total spectrum \citep[e.g.][]{Tanimoto25}. Consequently, the intrinsic \fexxvi\ absorption doublet signature  can be noticeably altered (if not significantly), leading to a spectral deviation from 1:2 line ratio (e.g. neutron star dippers and GX~13+1). 
%
%
While  beyond the scope of the present work, we plan to carry out \fexxvi\ doublet simulations in our future work using multi-dimentional Monte Carlo calculations with {\tt MONACO} by including scattered emission \citep[e.g.][]{Tanimoto25} in an effort to further examine the role of scattered X-rays  in the current framework of the disk wind model. 


Although emission lines are often observed to be very weak or absent in the observed wind spectra, high S/N  data 
occasionally implies a hint of  P-Cygni-like profile particularly more prominent in higher energy bandpass \citep[e.g., Fe K band;][]{Miller15,Tsujimoto25}, which is also intimately related to \fexxvi\ doublet signature (including obscured sources such as GRS1915+105, V4641 and V404 Cyg). It is thus important to investigate a correlation between the two in a comprehensive manner in a future work.   
Nonetheless, more microcalorimeter observations of disk winds will reliably unveil otherwise enigmatic \fexxvi\ spectral shape and its diversity.


\begin{acknowledgments}

The authors thank an anonymous referee for insightful comments to improve the manuscript. We are grateful to the {\it XRISM} team members who have inspired an idea and thoughts to motivate the present work. This work is supported in part by 
{\it XRISM} Guest Scientist (XGS) grant with NASA (80NSSC23K1021) and the Department of Physics and Astronomy of the College of Science and Mathematics at James Madison University (K.F.). The present research are partly supported by the Kagoshima University postdoctoral research program (KU-DREAM) (A.T.). F.T. acknowledges funding from the European Union - Next Generation EU, PRIN/MUR 2022 (2022K9N5B4).

\end{acknowledgments}

\begin{contribution}

K.F. designed and initiated the project and  all authors contributed to explanation and discussion, followed by preparing the manuscript.


\end{contribution}

%
\facilities{XRISM(Resolve), Chandra(HETGS) }

\software{HEASOFT/XSTAR}


%
%
%


\bibliography{FeXXVI}{}
\bibliographystyle{aasjournalv7}

\end{document}